\documentstyle[12pt]{article}
\begin{document}

\begin{center}
{\Large\bf Fermion mass gap in the loop representation 
of quantum gravity}  

\vspace{.5cm}
{\bf
Merced Montesinos--Vel\'asquez$^a$\footnote{e-mail:merced@fis.cinvestav.mx},
Hugo A. Morales--T\'ecotl$^b$\footnote{e-mail: hugo@xanum.uam.mx} \\
and  Tonatiuh Matos$^{a,c}$\footnote{e-mail: tmatos@fis.cinvestav.mx}\\
}
$^{a}$Departamento de F\'{\i}sica, Centro de Investigaci\'on y de Estudios
Avanzados \\ del I.P.N., P.O. Box 14--740, M\'exico D.F., 07000, MEXICO.\\
$^b$Departamento de F\'{\i}sica, Universidad Aut\'onoma
Metropolitana--Iztapalapa,\\ P.O. Box 55-534, M\'exico D.F., 09340, MEXICO.\\
$^c$Instituto de F\'{\i}sica y Matem\'aticas, Universidad Michoacana de San
Nicol\'as \\ de Hidalgo, P.O. Box 2-82, 58040 Morelia, Michoac\'an, MEXICO.

\end{center}

\begin{abstract}
An essential step towards the identification of a fermion mass generation 
mechanism at Planck scale is to analyse massive fermions  
in a given quantum gravity framework.
In this letter the two mass terms entering the Hamiltonian constraint for the 
Einstein-Majorana system are studied in the loop representation of 
quantum gravity and fermions. One resembles a bare mass gap because 
it is not zero  for states with zero 
(fermion) kinetic energy as opposite to the other that is interpreted 
as `dressing' the mass. The former contribution originates from 
(at least) triple intersections of the loop states acted on whilst the latter
is traced back to every couple of coinciding end points, where fermions sit. 
Thus, fermion mass terms get encoded in the combinatorics of loop states.
At last the possibility is discussed  of relating fermion masses to 
the topology of space.

\end{abstract}


\baselineskip=20pt

The physics at Planck length
$\ell_P:=\sqrt{\frac{G\hbar}{c^3}}=1.6\times 10^{-33}$cm,
where a quantum notion of spacetime is called for is acquiring 
deeper significance due to a number of new 
results. Amongst the most striking we find: i) a definition of a 
gravitational Hamiltonian \cite{clock1} 
and more recently, a skein-relation interpretation of the Hamiltonian 
constraint \cite{GambiniP1},
ii) a determination of area and volume spectra  \cite{volume} and 
iii) an insight on the origin of black hole entropy \cite{hole}.
All of them were obtained within a non perturbative approach to
quantum gravity \cite{Ashtekar,RovelliS,Gambini} and they  
present discrete and combinatorial features 
that seem to encode fundamental quantum aspects of  
spacetime at $\ell_P$.
For instance, by coupling a clock scalar field to gravity
a hamiltonian was built up in \cite{clock1}
that evolves the gravitational field itself.
The action of this hamiltonian on loop states is concentrated on intersection
points of the loops. More recently the Hamiltonian constraint 
was interpreted as  a skein relation when acting on the space  of 
knots \cite{GambiniP1}; hereby knot polynomials satisfying the skein relation
solve the full quantum Einstein equations.
In \cite{volume} it was shown that area and volume 
operators can be defined at the quantum level. Their spectra are discrete and
related to the way intersections occur between loops 
and the surface whose area should be 
determined or among loops inside the region whose volume is under study.
This former notion of area is further exploited in relation with the 
horizon of 
a black hole in \cite{hole}. Hence a value for its entropy 
can be estimated thereby that agrees with the standard 
proportionality between entropy and horizon area.
It should be stressed that each of these results was obtained after a 
suitable regularization procedure to make them well defined;
the resulting operators are finite, diffeomorphisms invariant
and (regularization-) background independent.

Now it is natural to wonder how compatible is the continuous picture of space 
we are used to with the above discreteness.
It turns out that smooth space can be thought of as a 
large length limit of certain loopy states or weaves \cite{weaves}. 
Moreover the notion of gravitons also emerges here:  they are 
associated to  
embroideries on weave states \cite{Junichi}.
Appealing as this idea is it cannot describe nature as a whole;
one must learn first what the notion of matter is, if any, consistently with
the above discrete picture. Indeed such a consistency must be looked for 
in any given quantum gravity scenario.

It has been proposed in the past that wormholes at the Planck scale might 
behave as charged particles \cite{MisnerWheeler} and that quantum gravity
states could have half integral angular momentum, when the space three 
manifold has non trivial topology \cite{FriedmanSorkin}. Also,
by following  
the path integral approach to quantum gravity,
it was realized that one might include the contribution 
of non orientable spacetimes 
(non orientable foam) in the correspondent amplitude. This together
with CP invariance could produce an effective mass for the otherwise massless
fermionic fields living on such spacetimes \cite{Parker}. 
The standard model of electroweak interactions
has been studied along similar lines by
taking a random Planck lattice as an effective theory 
coming from a Planck-scale foam spacetime \cite{Preparata}. 

In the loop representation of non perturbative quantum gravity
some steps have been given to unravel the notion of matter.
Coupled electromagnetic field and gravity were considered 
in \cite{GambiniP2} as a simple unified description of gravitational 
and electromagnetic interactions. 
Physical states were found parametrized by two loops each of which
carries information on both gravity and electromagnetism
-- the Chern-Simons functional and Jones Polynomial playing a role in
the analysis.  In ref. \cite{moralesR}
massless spin-$\frac{1}{2}$ fields and gravity were studied.
A (clock-) scalar field was coupled to gravity and a Hamiltonian
evolving both fermion and gravity fields was introduced. The fermionic
contribution gets concentrated at the end points of the curves (where  
fermions sit) of the loop states acted upon. More recently in \cite{Krasnov}
a kinematical analysis was developed for the Einstein-Maxwell-Dirac
theory. 

Crucial to the notion of matter is the understanding of the origin of fermion
masses given the unsatisfactory status in this regard of the standard model
of electroweak interactions \cite{Preparata}. To do so a 
compulsory  step is
to find the analogue of the term $m\overline{\Psi(x)}\Psi(x)$  that 
reveals the fermion mass in the lagrangian form of field theory. 
In this letter an analysis is given
of the mass of a spin-$\frac{1}{2}$ field of Majorana 
type coupled to gravity, using the loop representation for canonical quantum 
fermions and gravity of \cite{moralesR}. 
Specifically, we study the features inherited from the   
discreteness and combinatorial aspects appearing in such an approach.

To begin with lets recall the outcome of the canonical analysis for the 
Einstein-Majorana (EM) system using  Ashtekar variables. There are three
first class constraints namely the Gauss, vector and Hamiltonian ones
\cite{Ashtekar2}
\begin{eqnarray}
{ {\cal G}}_{AB} 
& := & -{\cal D}_a{\widetilde \sigma}^a\,_{AB} - \eta_{(A} {\widetilde
\theta}_{B)},\quad  
{ {\cal V}}_a 
:=  {\widetilde \sigma}^b\,_{AB} F_{ab}\,^{BA}
-{\widetilde \theta}_A{\cal D}_a\eta^A\quad {\rm and}\nonumber\\ 
{{ {\cal H}}} 
& := &
-\frac{1}{2} {\widetilde \sigma}^a\,_A\,^C\,
{\widetilde \sigma}^b\,_C\,^B\,F_{abB}\,^A 
- {\widetilde \sigma}^a\,_A\,^B
{\widetilde \theta}_B {\cal D}_a \eta^A\nonumber + m
\left((\widetilde{\sigma})^2 
\eta_A\eta^A-\frac{1}{4} {\widetilde \theta}^A
{\widetilde\theta}_A\right)\nonumber \\
& \equiv & {{\cal H}}_{\rm Einstein} 
+ {{\cal H}}_{\rm Weyl} 
+ {{\cal M}}_1
+ {{\cal M}}_2
\label{constraintEM}\,, 
\end{eqnarray}
$A_a\,^{AB}(x), \eta^A(x)$ being the configuration variables and 
${\widetilde \sigma}^a\,_{AB}(x), {\widetilde \theta}_A(x)$ the corresponding
canonical momenta\footnote{Notice that the two fermionic mass terms 
are non vanishing because the {\em two}-spinor field $\eta^A(x)$ 
is {\em Grassmann} valued (e.g. $\eta_A\eta^A=-2\eta^0\eta^1$); 
as opposed to the incorrect remark in \cite{Krasnov}.}.
Here the Majorana spin-$\frac{1}{2}$ field contribution to 
${ {\cal H}}$ consists of the last three terms in
(\ref{constraintEM}) of which ${{\cal M}}_1$
and ${{\cal M}}_2$ are related to mass.
To proceed to the quantum theory one has to
solve the problem of constructing their (regularized) quantum operator 
version. This is achieved by adopting loop variables, where the Gauss law
is automatically fulfilled, following \cite{clock1,moralesR}.
For spin-$\frac12$ and gravity loop variables were built up in 
\cite{moralesR} as 
\begin{eqnarray}
X[\alpha]& := &
\psi^A(\alpha_i)U_A\,^B[\alpha]\psi_B(\alpha_f), 
\quad
Y[\alpha]  :=  {\widetilde
\pi}^A(\alpha_i)U_A\,^B[\alpha]\psi_B(\alpha_f),\nonumber\\ 
Y^a [\alpha](s) & := & {\widetilde
\pi}^A(\alpha_i)U_A\,^B[\alpha](0,s){\widetilde
\sigma}^a_B\,^C(\alpha(s))U_C\,^D[\alpha](s,1)\psi_D(\alpha_f)\, .
\end{eqnarray}
Among their properties it is worth mentioning 
the fermionic (Grassmann) identity: if ($\alpha, \beta, \gamma$ are open 
curves) $\alpha_i =\beta_i =\gamma_i$ then 
$X[\alpha]\, X[\beta] \,X[\gamma] = 0$. No three fermions can concide
at the same point simultaneously.
The loop variable $Y^a[\alpha](s)$  was used to define 
the kinetic fermion term of the Hamiltonian constraint of the Einstein-Weyl 
theory \cite{moralesR}. 
Next the first mass term ${{\cal M}}_1$ is 
translated into loop variables. 
Consider a closed loop $\gamma$ with three gravitational hands inserted 
in it and an open loop $\alpha$ with the fermion field $\eta(x)$ placed at 
its ends. That is to say 
\begin{equation}
V^{abc}[\gamma,\alpha]:= T^{abc}[\gamma](s,t,r)\, X[\alpha],\label{manita}
\end{equation}
where 
$T^{abc}[\gamma](s,t,r)  :=  \mbox{Tr} \left\{
{\widetilde\sigma}^a(\gamma(s))U_{\gamma}(s,t)
{\widetilde\sigma}^b(\gamma(t))U_{\gamma}(t,r)
{\widetilde\sigma}^c(\gamma(r))U_{\gamma}(r,s)\right\}$
is the loop variable used in the construction of the 
volume operator \cite{volume} and 
$X[\alpha]$ was given above.
It is straightforward to show that
when the two loops shrink down to a {\em common} 
point $x$ we have the local quantity
\begin{eqnarray}
{{\cal M}}_1 =
\left(\frac{m}{3\sqrt{2}}\right)
\lim_{\gamma,\alpha\rightarrow x} \eta_{abc} V^{abc}[\gamma,\alpha]
= m (\sigma)^2\eta_A\eta^A.
\end{eqnarray}

The construction  becomes more transparent if
Eq.(\ref{manita}) is rewritten as follows.
Then use is made of the fundamental spinor
identity $\epsilon_{AB}\epsilon_{CD}+\epsilon_{AC}\epsilon_{DB}=
\epsilon_{AD}\epsilon_{CB}$ inserted at the intersection point of 
the two loops. The result is the difference of further loop 
variables 
\begin{eqnarray}
T^{abc}[\gamma](s,t,r)\,X[\alpha] & = &
N^{abc} [\alpha\cdot\gamma](s^{\ast}, t^{\ast}, r^{\ast}) \nonumber\\
& & - N^{cba} [\alpha\cdot\gamma^{-1}]
(1-s^{\ast},1-t^{\ast},1- r^{\ast})\\
N^{abc}[\alpha\cdot\gamma](s^{\ast}, t^{\ast}, r^{\ast}) & := &
\mbox{Tr} \left\{ \,
\eta(\alpha_i)U[\alpha](0,p)U[\gamma](0,s){\widetilde\sigma}^a
(\gamma(s))U[\gamma]
(s,t){\widetilde\sigma}^b(\gamma(t))\cdot\right.\nonumber\\
& & \left. U[\gamma](t,r){\widetilde\sigma}^c(\gamma(r))
U[\gamma](r,1)U[\alpha](p,1)\eta(\alpha_f) \right\} , 
\end{eqnarray}
with $\alpha(p)=\gamma(0)=\gamma(1)=x$ and $s^{\ast},t^{\ast},r^{\ast}$ are 
the values of the parameter of $\alpha\cdot\gamma$, where the gravitational 
hands are inserted. Here 
Tr$\left\{ \psi {\cal  O}^{(1)}\dots {\cal O}^{(n)} \psi \right\}:= 
\psi^{A_1} {\cal O}_{A_1}^{(1){}A_2} \dots {\cal O}_{A_n}^{(n){}A_{n+1}} 
\psi_{A_{n+1}}$.

Regarding ${{\cal M}}_2$, it can be 
expressed as 
\begin{equation}
{{\cal M}}_2 := 
-\left(\frac{m}{4}\right) \lim_{\alpha\rightarrow x} Z[\alpha] \quad
{\rm with} \quad Z[\alpha] := 
{\widetilde\theta}^A(\alpha_i)\,U_A\,^B[\alpha]\,
{\widetilde\theta}_B(\alpha_f),
\label{MTEM2}
\end{equation}
when $\alpha$ shrinks down to the point $x$. 

Based on the loop transform of \cite{moralesR} 
the action of the (non regularized) operators 
(\ref{manita}) and (\ref{MTEM2}) can be defined on loop states as
\begin{eqnarray}
{\widehat Z}[\alpha]\,\Psi\, [\beta] 
& = & 
\delta ^3(\alpha_f,\beta_i)\,\delta
^3(\alpha_i,\beta_f)\,\Psi[\alpha\cdot\beta] \nonumber\\
& & + \delta ^3(\alpha_f,
\beta_f)\,\delta ^3(\alpha_i,\beta_i)\, \Psi
[\alpha\cdot\beta^{-1}], \label{zq}\\
{\widehat V}^{abc}[\gamma,\alpha] \Psi[\beta]
 &=&
\sum_{\mu=\pm 1}\sum^8_{j=1}\sum^6_{i=1}\left(\frac{1}{\sqrt{2}}\right)^3
\cdot \nonumber\\
& & \left(\Delta^a[\alpha\cdot\gamma^{\mu}(s^*),\beta]
\Delta^b[\alpha\cdot\gamma^{\mu}
(t^*),\beta]
\Delta^c[\alpha\cdot\gamma^{\mu}(r^*),\beta]\right)_i \cdot \nonumber\\
& & (-1)^{r_{ij}}{c_{ij}(-1)^{\frac12 (1-\mu)}}\Psi
[(\alpha\cdot\gamma^{\mu}\cdot\beta)_{ij}]\, ,\label{vq}\\
\Delta^a[\alpha\cdot\gamma^{\mu}(s^*),\beta]
&=& \frac{1}{2} \int_{0}^{1} du \dot{\beta}^{a}(u) 
     \delta^3(\alpha\cdot\gamma^{\mu}(s^*),\beta(u))\, . \nonumber
\end{eqnarray}
${\widehat V}^{abc}[\gamma,\alpha]$ produces 
sixteen multiloop states\footnote{Of these sixteen contributions, eight come  
from attaching the open loop $\beta$ to the 
loop $\alpha\cdot\gamma$ and the other eight by attaching the open 
loop $\beta$ to $\alpha\cdot\gamma^{-1}$.} 
(indexes $\mu$ and $j$) for each exclusive 
configuration labeled by $i$. In other words, 
$(\Delta^a[\alpha\cdot\gamma^{\mu}(s),\beta]
\Delta^a[\alpha\cdot\gamma^{\mu}(t),\beta]
\Delta^a[\alpha\cdot\gamma^{\mu}(r),\beta])_i$ represent the six 
different ways in which the open loop $\beta$ is attached to the open loop 
$\alpha\cdot\gamma^{\mu}$. $\Psi[(\alpha\cdot\gamma^{\mu}\cdot\beta)_{ij}]$ 
denote the multiloop states resulting from 
rerouting $\alpha\cdot\gamma^{\mu}$ 
and $\beta$. $r_{ij}$ is the number of orientation-reversed segments of 
any $\alpha\cdot\gamma^{\mu}$ or $\beta$-loop segments between intersections 
to get a consistent overall orientation while the parameter $c_{ij}$ is
such that $c_{ij}=-1$ if the multiloop
$\Psi[(\alpha\cdot\gamma^{\mu}\cdot\beta)_{ij}]$ has open component
starting at $\alpha_i$ and ending at $\beta_f$; otherwise
$c_{ij}=+1$.

To regularize the mass terms ${{\cal M}}_1$
and ${{\cal M}}_2$ amounts to regularize
(\ref{zq}) and (\ref{vq}). To this aim use is made of  
an auxiliary background flat metric and a prefered set of 
coordinates in which this metric is Euclidean.
Lets take the partition of the 3-dimensional fictitious space in 
cubes of sides $L$. 
The regularised quantum version of the mass operators 
is readily  performed by introducing \cite{clock1,moralesR}
\begin{equation}
\widehat{{\cal M}}:= \lim_{L,\xi\rightarrow 0} \sum_I L^3
\sqrt{-\widehat{\cal M}_{1\,I}^{L} - 
\widehat{\cal M}_{2\,I}^{L,\xi}} \label{M}\,.
\end{equation} 
Lets describe $\widehat{\cal M}_{1\,I}^{L}$ 
($\widehat{\cal M}_{2\,I}^{L,\xi}$ will be analysed below.)
\begin{eqnarray}
{\widehat {{{\cal M}}}}_{1\,I}^L & := & 
\frac{\sqrt{2} m}{8\cdot 3! L^6}
\int_{\partial I}d\sigma^2
\int_{\partial I}d\tau^2
\int_{\partial I}d\rho^2
(-1)^{r_a+r_b+r_c}\eta^{abc}n_a(\sigma)n_b(\tau)n_c(\rho)
{\widehat V}^{abc}[\gamma,\alpha],\nonumber\\
& = & \frac{\sqrt{2} m}{8\cdot 3! L^6}
\int_{\partial I}d\sigma^2
\int_{\partial I}d\tau^2
\int_{\partial I}d\rho^2
(-1)^{r_a+r_b+r_c}\eta^{abc} n_a(\sigma)n_b(\tau)n_c(\rho)
\cdot\nonumber\\
&\quad & \quad\left\{{\widehat N}^{abc}
-{\widehat N}^{cba}\right\},
\label{m1r}
\end{eqnarray}
where one particular box has been considered (in the fictitious metric). 
$\partial I$ indicate its boundary, namely the union of the six faces of
the cube oriented outwards and $\gamma$ and $\alpha$ 
are closed and open, respectively, with a common point of intersection. 
The closed loop $\gamma$ has three gravitational hands
inserted lying on the boundary of the 
box at the point $\sigma,\tau,$ and $\rho$. The loop $\gamma$ is the 
triangle formed by three segments that connect the three points
$\sigma, \tau$ and $\rho$ and 
$\gamma(s)=\sigma, \gamma(t)=\tau, \gamma(r)=\rho$. The open loop has two 
fermions at its ends, $\eta(\alpha_i)$ and $\eta(\alpha_f)$, respectively. 
These fermions are in the box; $n_a$ is the normal one-form to the box's 
boundary. No sum convention is applied to $r_a$ and $\eta^{abc}$; 
$r_a=0$ at the front and $r_a=1$ at the back of the boundary. 

The action of ${\widehat{{ {\cal M}}}}^{L}_{1\, I}$ 
on the loop states is as follows. 
The three surface integrals on the boundary of the box $I$ and the three line 
integrals along the loop $\beta$ that parametrize the loop state combine to 
give three numbers related to the intersections of the open loop with the 
boundary of the cube. The non vanishing contribution can be traced back 
to the intersection of the open loop 
$\beta$ simultaneously with three different faces of the cube. That is to say 
when the open loop $\beta$ has {\em at least} a triple point of intersection 
in the cube. 
The cube shrinks down to that point in the limit $L\rightarrow 0$. 
Hereby the gravitational hands are smeared on the boundary of the cube, 
for each 
permutation of them there are $8\cdot16$ terms which correspond to all the 
posibilities in which the hands can lie on the faces of the cube. Each
one of these terms has, in general, different weight because of two reasons: 
the first one is due to the orientation of the open loop $\beta$ when it 
intersects one of the three faces of the cube. The second one comes from 
the factor $+1$ or $-1$ depending on whether the hands lie at the front or 
the back faces. Note that of these $8\cdot 16$ terms 
only one particular permutation of the hands enters once: $16$  of 
them contribute for a specific loop $\beta$ since each 
hand intersects the open loop once. By taking into account the six 
permutations of the hands there are, for a specific loop, $16\cdot 3!$ terms 
and for a general situation $8\cdot 16\cdot 3!$ terms. 

It is important to 
mention that the prescription given above depends in a way on the open 
loop $\beta$ labeling the loop state. More precisely, one needs to know 
some topological information (intersections and end points)
of $\beta$ in order to calculate its specific contribution. 
Observe that the prefactor in (\ref{m1r}) is finite in the $L\rightarrow 0$
limit because the surface integrals produce a factor $L^6$ that
cancels out the one in the denominator. Hence one gets a finite action of
the operator. This is analogous to the case of the volume operator
\cite{volume}.

Due to the fact that 
${\widehat {{ {\cal M}}}}^{L}_{1\, I}$ 
has only gravitational hands it can even ``see" loop states without 
fermionic excitations, namely loop states parametrized by closed loops with 
at least a triple point of intersection. This is precisely the difference
with the kinetic fermion term \cite{moralesR} 
(i.e. ${{\cal H}}_{\rm Weyl}$ above) which is ``blind" to 
this type of loop states (it yields zero on such states.) Hence the 
interpretation here proposed for ${\widehat{\cal M}}_1$ is as
forming the {\em gap fermion mass}. This is the major result presented here.

For the case of $\widehat{\cal M}_2$ 
and hence ${\widehat Z}[\alpha]$ consider 
$(\alpha_{\vec x,\vec y})(s)$ a  
straight line (in the background metric) that starts at 
$\vec x$ an points in the $\vec y$ direction
\begin{eqnarray}
(\alpha_{\vec x,\vec y})(s)= \vec x + s\vec y,
\quad (\alpha_{\vec x,\vec y})(0)=\vec x,
\quad (\alpha_{\vec x,\vec y})(1)=\vec x +\vec y\, .
\end{eqnarray}
Then define the following
\begin{eqnarray}
{\widehat {{{\cal M}}} }^{L,\xi}_{2\,\,I} 
& := & \frac{1}{L^3} \int_I d ^3x\,
{\widehat{{{\cal M}}}}^{\xi}_{2} (\vec x),\\ 
{\widehat {{{\cal M}}}}^{\xi}_{2}(\vec x) 
& := & \frac{\frac{-m_D}{4}}{\frac{4}{3} \pi\xi^3}\int d ^3y \,
\theta (\xi-\mid \vec y \mid)\,{\widehat Z}[\alpha_{\vec x ,\vec y}],
\end{eqnarray}
where $\xi<L$ and $\theta$ is the step function.

Now if the end points of the open loop $\beta$, $\beta_i$ and $\beta_f$,
coincide with the end points of the open loop $\alpha$ inside the ball 
centered at $\vec x$ and with radius $\xi$ one has
\begin{eqnarray}
\lim_{L,\xi\rightarrow 0}\sum_{I}L^3\sqrt{{-\widehat {{
{\cal M}}}}^{L,\xi}_{2\, I}}
\Psi\,[\beta]
 & = & 
\lim_{L\rightarrow 0,
\xi\rightarrow 0}
\sqrt{\frac{\frac{m}{4}}{\frac43\pi\xi^3}\,L^3}({\widehat{\cal F}}_{\beta_i}
+{\widehat{\cal F}}_{\beta_f})^\frac12 \,\Psi\,[\beta],\label{dedo}\\
{\widehat{\cal F}}_e\,\Psi\,[\beta]& = &\left\{ \begin{array}{ll}  \Psi
\,[\alpha_{\vec x , \beta_i -\vec x}\cdot\beta] & \mbox{if $e=\beta_f$}\\ \Psi
\,[\alpha_{\vec x , \beta_f -\vec x}\cdot\beta^{-1}] &  \mbox{if
$e=\beta_i$}\,.
\end{array} \right.
\end{eqnarray}
The regularization parameters can be chosen as
\footnote{This election corrects the one in \cite{moralesR}.}            
$L(\epsilon)  =  b\, \epsilon ,\quad{\rm and }\quad 
\xi (\epsilon) =  b\,\sin{\epsilon}$,
where $b$ is an arbitrary length. Remarkably the prefactor in 
(\ref{dedo}) is finite and given by
\begin{eqnarray}
\lim_{\epsilon\rightarrow 0} C(L(\epsilon),\xi(\epsilon)) 
=\sqrt{\frac{\frac{m}{4}}{\frac43\pi\xi^3}L^3} = \sqrt{\frac{3m}{8\pi}}.
\end{eqnarray}
In this way, as opposed to $\widehat {\cal M}_1$, only loops with pairs of
coinciding point-like fermion 
excitations have contribution to  
$\widehat {\cal M}_2$, 
which is quadratic in the 
fermion momentum variables. In this respect is rather similar to the 
kinetic energy fermion contribution to the Hamiltonian constraint
${{\cal H}}_{\rm Weyl}$. However since anyhow it 
modifies the fermion mass a ``dressing" interpretation seems more 
appropriate.

In summary the Majorana type mass for fermions has been studied in the
loop representation of non perturbative quantum gravity and fermions.
There are two contributions one of which resembles a mass gap whereas
the other seems to dress the corresponding mass. The former is non zero 
for loop states even lacking fermion excitations and containting at least
triple intersections. The latter requires the presence of coinciding 
pairs of end points characterising the loop states.
Setting the Einstein and kinetic fermion terms to zero, 
the Majorana mass operator turns out to be 
\begin{eqnarray}
{\widehat M}_{\rm Majorana} = \sum_{\rm i,e} 
\sqrt{ {\widehat {{ {\cal M}}}}_{1}^{({\rm i})} +
{\widehat {{{\cal M}}}}_2^{({\rm e})} } \label{MM}
\end{eqnarray}
where the sum runs over (at least triple) intersections i
of the loop states (with and 
without fermionic excitations) and end points e for open loops with pairs 
of coinciding point-like fermionic excitations. The  limit in
which the regularization parameters go to zero is understood on the r.h.s. 
of Eq.(\ref{MM}). Some further comments are in order.


{\em Topology of space}.
Recently Smolin put forward the equivalence between minimalist quantum 
wormholes (i.e. identifying pairs of space points) 
without matter and quantum Einstein-Weyl theory  expressed in 
loop variables \cite{awh}. In this picture, the fermionic character 
of the Weyl field being associated to the antisymetrization of the mouths 
of the wormholes. In this way the fermionic matter gets  
encoded in the topological properties of the space.   
Also, 
non minimalist wormholes (that is smooth manifolds) could be 
considered having the results of the minimalist ones as their low energy 
limit \cite{awh}. This in turn sugests a scenario of the kind a Weyl field
living on a space foam as the one considered by Friedman et al \cite{Parker}
that yields an effective theory in which the fermion field becomes massive!
Nevertheless, the analysis of \cite{Parker} relies on a perturbative approach
which it is better to avoid in the loop representation. A way out consists  
in following Smolin's strategy of studying the equivalence of the 
Einstein-Majorana theory, as given in the present work, to the non 
minimalist quantum wormholes. Further work is needed to settle down 
the issue of generating fermion masses out of the topology of the space
in the context of non perturbative quantum gravity.

A mechanism of mass generation would allow to calculate the values of the
masses of course but realistic values, i.e. related to nature, might well 
only come from the incorporation with the other non gravitational fundamental 
interactions. This possibility was left open but some steps are in 
progress \cite{masa2} along the lines presented here combined 
with those of \cite{Preparata}. Also, a massive Dirac field is currently
under study.

{\em Reality conditions and spin networks}.
Relying  on Ashtekar approach for gravity and spin-$\frac12$ fields 
involves two complex local degrees of freedom for the gravitational field 
unless reality
conditions are suplemented \cite{David}. Of course the question remains open
that such reality conditions will single out the correct inner product at the
quantum level. Nevertheless, the present analysis is expected to be robust
enough to encompass real variables along the lines of \cite{Thiemann}.
This will be possible after extending the spin-network framework to include
spin-$\frac12$ fields, as in \cite{Pietri}.

{\em Acknowledgements.}
Partial support from CONACyT Grants 3141P-E9607 and
E120-2639 (joint with CONICyT-Chile) is acknowledged. MMV
has been supported by CONACyT Reg. No. 91825. It is a pleasure
to thank C. Rovelli for his encouragement in the preparation of this work.

\newpage

\end{document}